\documentclass[aps,prl,twocolumn,groupedaddress,showpacs]{revtex4}
\usepackage{amsmath}
\usepackage{amssymb}
\usepackage{amsopn}
\usepackage{graphics}
\usepackage{graphicx}
\usepackage{color}

\newcommand{\dd}[0]{\mathrm{d}}
\newcommand{\bb}[0]{\begin{eqnarray}}
\newcommand{\ee}[0]{\end{eqnarray}}
\newcommand{\nn}{\nonumber}

\newcommand{\sn}[1]{\sigma_{#1}}
\newcommand{\PF}[0]{\mathcal{Z}}

\newcommand{\prodd}[2]{\overrightarrow{\prod_{#1}^{#2}}}
\newcommand{\prodg}[2]{\overleftarrow{\prod_{#1}^{#2}}}

\newcommand{\Cst}{u_0}
\newcommand{\opx}[1]{\mathcal{O}_{#1}}
\newcommand{\opy}[1]{\mathcal{P}_{#1}}
\newcommand{\baropx}[1]{\bar{\mathcal{O}}_{#1}}
\newcommand{\baropy}[1]{\bar{\mathcal{P}}_{#1}}
\newcommand{\coefk}[1]{\alpha_{#1}}
\newcommand{\sgn}[1]{\sigma({#1})}
\newcommand{\pol}{u}
\newcommand{\mass}{m}
\newcommand{\ch}{\textrm{ch}}
\newcommand{\sh}{\textrm{sh}}

\DeclareMathOperator*{\TR}{\mathfrak{Tr}}

\begin{document}
\title{On second-order critical lines of spin-S Ising models
in a splitting field with Grassmann techniques}
\author{J.-Y. Fortin}
\email{fortin@lpt1.u-strasbg.fr}
\affiliation{Laboratoire de Physique Th\'eorique, CNRS/UMR 7085 and
Universit\'e Louis Pasteur, 67084 Strasbourg Cedex, France}
\author{M. Clusel}
\email{maxime.clusel@nyu.edu}
\affiliation{Department of Physics and Center for Soft Matter Research,
New-York University, 4 Washington Place, New-York NY 10003, USA}
\date{\today}

\begin{abstract}
We propose a method to study the second-order critical lines of classical spin-$S$ Ising models on two-dimensional lattices in a crystal or splitting field, using an exact expression for the bare mass of the underlying field theory. Introducing a set of anticommuting variables to represent the partition function, we derive an exact and compact expression for the bare mass of the model including all local multi-fermions interactions. By extension of the Ising and Blume-Capel models, we extract the free energy singularities in the low momentum limit corresponding to a vanishing bare mass. The loci of these singularities define the critical lines depending on the spin $S$, in good agreement with previous numerical estimations. This scheme appears to be general enough to be applied in a variety of classical Hamiltonians.
\end{abstract}

\pacs{02.70.Rr,05.70.Jk,64.60.De,68.35.Rh}
\maketitle

Classical or quantum spin models play a central role in the development of statistical physics, as they allow for testing fundamental principles on model systems that are simple enough to have a complete mathematical description. This is particularly the case for the Ising model, which still serves as a toy model to develop new techniques both in terms of analytical methods (series expansions, renormalization) and numerical algorithms (Metropolis, Swendsen-Wang). The fact that an analytic solution is available in the two-dimensional case provides a reference point for the further understanding of critical phenomena. While the case of the spin $S=1/2$ is well understood, much less is known about the extension for spins with higher values. In the early 70's, Fox and Guttmann successfully developed  low temperature series expansions, allowing for an estimation of the critical temperatures and set of critical exponents for the Ising model with spin $S=1$ and $3/2$, in two and three dimensions \cite{Fox73}. This original work has later been extended to higher spin values \cite{Enting94,Jensen96}. High temperature series expansions have also been used in three dimensions \cite{Saul75,Camp75} and two dimensions \cite{Butera03}, and then extended to test universality and hyperscaling, as well as spin-spin correlation functions \cite{Butera02,Butera03}. Another path of investigation is provided by the Fortuin-Kasteleyn transformation \cite{Fort72}, which gives a fruitful link between Ising models and percolation problems \cite{Fort02}, leading to interesting results in the continuous spin limit \cite{Bial00}. Alternative approximation schemes have been proposed to treat the general spin case, such as Ising spin decomposition of the general spin-S model \cite{Horiguchi95}, an effective mean-field theory based on cumulant expansion \cite{Li01} and Husimi tree calculation \cite{Monroe02}. However despite this activity, very few exact results have been obtained on the general spin-S Ising models, even in two dimensions. Results are available for the special case $S=1$ in particular regions of the phase diagram \cite{Horiguchi86,Wu86,Shankar86,Rosengren89}, and for the case $S=3/2$ \cite{Izma94}. Recently, we proposed an alternative approach to the Blume-Capel model where $S=1$ \cite{Clusel08}, using the representation of the partition function with Grassmann variables \cite{Plechko85} to determine the critical line of this model. In this article, we extend this method to study the general spin-$S$ model. The first step is to expand the partition function as a product of spin polynomials where neighboring spins are coupled. We then introduce a set of Grassmann variables \cite{berezin} to decouple the spins. The price to pay for this is the lost of commutativity. However, using specific symmetries, the sum over the spin degrees of freedom can be performed exactly, leading to the expression of the partition function as a path-integral over a fermionic action on a lattice. Taking the thermodynamical limit of the action allows us to identify the bare mass of the system. Assuming that free energy singularities correspond to the vanishing mass, we obtain an excellent approximate location of the critical points.\\ 
We consider now the general Hamiltonian on a two-dimensional (2D) lattice of size $L\times L$

\begin{eqnarray}
\label{ham1}
\begin{split}
H = -\sum_{m=1}^{L}\sum_{n=1}^{L}\Big[J_1 \sn{mn}\sn{m+1n}
+J_2 \sn{mn}\sn{mn+1}\Big] 
\\
+\Delta_0\sum_{m=1}^{L}
\sum_{n=1}^{L}\sn{mn}^{2},
\end{split}
\end{eqnarray}
where $J_{1,2}$ are Ising coupling constants and $\Delta_0$ a splitting 
crystal field that favors small spin values. It can also represent a chemical potential in the Blume-Capel model. In the Ising case, $\sn{mn}^2=1$, and therefore $\Delta_0$ does not play any role. This crystal field term can be replaced by any
potential $V(\sn{mn}^2)$ depending on the square of the local spin.
 The spins $\sn{mn}$ can take $2S+1$ values with $\sn{mn}=-S,-S+1,\cdots,S$.
We assume here for simplicity that $J_1=J_2=1$ and $K=\beta J_{1,2}=1/k_BT$ the
inverse of temperature. The partition function, which represents the sum over all 
possible spin configurations $\PF=\TR_{\{\sn{}\}}\exp(-\beta H)$, contains products of the Boltzmann weights $\exp(K\sn{}\sn{}')$ (where $\sn{}$ and $\sn{}'$ are neighboring spins) which take $q+1=S(S+1)+1$ different values if $S$ is an integer, and $q+1=(S+1/2)(S+3/2)$ values if $S$ is half-integer. 
Since there are $q+1$ possible values for each Boltzmann weight $\exp(K\sn{}\sn{}')$, we can project each of them onto a polynomial function of degree $q$ in variable $\sn{}\sn{}'$:
\bb\label{Boltz}
\exp(K\sn{}\sn{}')&=&\sum_{k=0}^{q}\pol_k(\sn{}\sn{}')^k
=\Cst\prod_{\alpha=1}^{q}(1+x_{\alpha}\sn{}\sn{}'),
\ee
where the $q+1$ constants $\pol_k$ are determined by solving the linear 
system of $q+1$ equations satisfied by the above relation. Note that the demonstration below holds for any function of $\sn{}\sn{}'$ in the Boltzmann factor, in the case where the Hamiltonian (\ref{ham1}) includes quadrupolar interactions $(\sn{}\sn{}')^2$ for example \cite{Izma94}. Only the coefficients $\pol_k$
are different.
In the Ising case, $S=1/2$ and $q=1$, we have $\exp(K\sn{}\sn{}')=\ch(K/4)+4\sh(K/4)\sn{}\sn{}'$, and $\pol_0=\ch(K/4)$, $\pol_1=4\sh(K/4)$.
In the Blume-Capel model $q=2$ and it is easy to show that $\pol_0=1$, $\pol_1=\sh(K)$ and $\pol_2=\ch(K)-1$. For $S$ integer we always have $\pol_0=1$, and from Eq. (\ref{Boltz}),
$\pol_{1\le k\le q}=\Cst\sum_{\alpha_1<\alpha_2<\cdots<\alpha_k}x_{\alpha_1}x_{\alpha_2}\cdots x_{\alpha_k}$. We also set $\pol_{k\ge q+1}=0.$\\

Let us now introduce $q$ pairs of Grassmann variables  $(a_{mn}^{\alpha},\bar{a}_{mn}^{\alpha})$ at each site for the horizontal direction and $q$ additional pairs $(b_{mn}^{\alpha},\bar{b}_{mn}^{\alpha})$ for the vertical direction \cite{berezin,Plechko85}. Here $\alpha$ takes the values 
$1,\dots q$. In total, there are $4q$ Grassmann variables on each site
of the lattice. The Ising model is usually represented by two pairs of Grassmann variables per site which can be reduced afterwards to one pair \cite{Plechko85}. We then use the following integral representation for each couple of terms 
$(1+x_{\alpha}\sn{mn}\sn{m+1n})$ and $(1+x_{\alpha}\sn{mn}\sn{mn+1})$ that appear in Boltzmann weights Eq. (\ref{Boltz}):
\bb \nn
& &1+x_{\alpha}\sn{mn}\sn{m+1n}=
\\ \nn
& &\int d\bar{a}_{mn}^{\alpha}da_{mn}^{\alpha}
{\rm e}^{a_{mn}^{\alpha}\bar{a}_{mn}^{\alpha}}(1+a_{mn}^{\alpha}\sn{mn})
(1+x_{\alpha}\bar{a}_{mn}^{\alpha}\sn{m+1n}),
\\ 
& &1+x_{\alpha}\sn{mn}\sn{mn+1}=
\\  \nn
& &\int d\bar{b}_{mn}^{\alpha}db_{mn}^{\alpha}
{\rm e}^{b_{mn}^{\alpha}\bar{b}_{mn}^{\alpha}}(1+b_{mn}^{\alpha}\sn{mn})
(1+x_{\alpha}\bar{b}_{mn}^{\alpha}\sn{mn+1}).
\ee
From these expressions, we introduce the link factors $A_{mn}^{\alpha}=1+a_{mn}^{\alpha}\sn{mn}$, $\bar{
A}_{m+1n}^{\alpha}=1+x_{\alpha}\bar{a}_{mn}^{\alpha}\sn{m+1n}$,
$B_{mn}^{\alpha}=1+b_{mn}^{\alpha}\sn{mn}$, and $\bar{
B}_{mn+1}^{\alpha}=1+x_{\alpha}\bar{b}_{mn}^{\alpha}\sn{mn+1}$. Then
the partition function can be written as
\begin{multline}\nn
\frac{\PF}{\Cst^{2L^2}}=\TR_{\{\sn{}\}}\int
\prod_{mn,\alpha}d\bar a_{mn}^{\alpha}da_{mn}^{\alpha}
d\bar b_{mn}^{\alpha}db_{mn}^{\alpha}
e^{a_{mn}^{\alpha}\bar a_{mn}^{\alpha}+b_{mn}^{\alpha}\bar b_{mn}^{\alpha}}
\\ \nn
\times
\prod_{mn}
e^{\Delta\sn{mn}^2}
\Big [
\prod_{\alpha}
(A_{mn}^{\alpha}\bar A_{m+1n}^{\alpha})
\prod_{\beta}
(B_{mn}^{\beta}\bar B_{mn+1}^{\beta})
\Big ],
\end{multline}
where $\Delta=-\beta\Delta_0$. Notice that inside the integral symbol, the pairs of link factors in brackets $(A_{mn}^{\alpha}\bar{A}_{m+1n}^{\alpha})$ and $(B_{mn}^{\alpha}\bar{B}_{mn+1}^{\alpha})$ can move freely. In particular, we can rearrange the products over 
$\alpha$ and put together link factors having the same site indices $(m,n)$ using the {\it mirror} ordering symmetry \cite{Plechko85}
\bb\nn
\prod_{\alpha=1}^{q}
(A_{mn}^{\alpha}\bar{A}_{m+1n}^{\alpha})=
\left(
\prodd{\alpha=1}{q}A_{mn}^{\alpha}\right)\left(\prodg{\alpha=1}{q}\bar{A}_{m+1n}^{\alpha}
\right),
\ee
where the arrows indicate that the product is ordered, i.e. increasing 
label $\alpha$ in the first product from left to the right and in the second
one from right to the left. For convenience, we set $\opx{mn}=\prodd{\alpha}{}A_{mn}^{\alpha}$, $\baropx{m+1n}=\prodg{\alpha}{}\bar{A}_{m+1n}^{\alpha}$ and $\opy{mn}=\prodd{\alpha}{}B_{mn}^{\alpha}$, $\baropy{mn+1}=\prodg{\alpha}{}\bar{B}_{mn+1}^{\alpha}$.
Then the partition function can be rewritten as

\bb\nn
\frac{\PF}{\Cst^{2L^2}}&=&\TR_{\{\sn{}\}}
\int
 \prod_{mn,\alpha}d\bar{a}_{mn}^{\alpha}da_{mn}^{\alpha}
 d\bar{b}_{mn}^{\alpha}db_{mn}^{\alpha}
 e^{a_{mn}^{\alpha}\bar{a}_{mn}^{\alpha}+b_{mn}^{\alpha}\bar{b}_{mn}^{\alpha}}
 \\ 
 & &
 \times
 \prod_{mn}
 e^{\Delta\sn{mn}^2}
 (\opx{mn}\baropx{m+1n})(\opy{mn}\baropy{mn+1}),
 \\ \nn
 &\equiv&
\TR_{\{\sn{},a,\bar{a},b,\bar{b}\}} \left[
\prod_{mn}e^{\Delta\sn{mn}^2}
(\opx{mn}\baropx{m+1n})(\opy{mn}\baropy{mn+1}).
\right]
\ee
At this stage, we use the {\it mirror} and {\it associative} symmetries which were applied to the Ising model \cite{Plechko85b} and which are still valid here to rearrange the operators $\opx{}$, $\baropx{}$, $\opy{}$ and $\baropy{}$. The computations are until now identical to the Ising case treated in references \cite{Plechko85b,Plechko2002}, in the sense that we obtain an expression of the partition function with a set of anticommuting operators we would like to rearrange in order to perform the sum over the individual spins. The only difference is that the previous operators $\opx{}$, $\baropx{}$, $\opy{}$ and $\baropy{}$ are in general more complicate functions of the $4q$ Grassmann variables coming from the decomposition given by relation (\ref{Boltz}).
In principle boundary terms should be treated separately in order to obtain the exact finite size partition function. Periodic boundary conditions can be treated rigorously for finite lattice  \cite{Plechko85,wu02,CF1} but this is inessential in the thermodynamical limit $L\rightarrow\infty$ we are interested in here. We consider instead free boundary conditions, leading to the exact expression :
\bb\nn 
\frac{\PF}{\Cst^{2L^2}}=
\TR_{\{\sn{},a,\bar{a},b,\bar{b}\}} \left[
\prodd{n=1}{L}
\Big (
\prodd{m=1}{L}
e^{\Delta\sn{mn}^2}
\Big (
\baropx{mn}\baropy{mn}\opx{mn}
\Big )
\prodg{m=1}{L}
\opy{mn}
\Big )
\right].
\ee
Under this form, the spins can individually be summed over from $\sn{Ln}$ to
$\sn{1n}$ for any given $n$. We introduce the following weights $W_{mn}$ which include all the dependence on the individual spin $\sn{mn}$ 
\bb\nn
W_{mn}=\sum_{\sn{mn}=\pm 1}
e^{\Delta\sn{mn}^2}
\baropx{mn}\baropy{mn}\opx{mn},
\opy{mn}
\\ \label{sum1}
\equiv
\sum_{\sn{mn}=\pm 1}
e^{\Delta\sn{mn}^2}
\prodd{\alpha=1}{4q}\Big (
1+c_{mn}^{\alpha}\sn{mn}
\Big ),
\ee
where we have defined the following sets of Grassmann variables
\begin{eqnarray} \nn
\label{defc}
c_{mn}^{\alpha}=\left\{
   \begin{array}{ll}
          x_{q-\alpha+1} \bar{a}_{m-1n}^{q-\alpha+1}& \text{if } \alpha=1,\cdots, q,\\ 
	  x_{2q-\alpha+1} \bar{b}_{mn-1}^{2q-\alpha+1}& \text{if } \alpha=q+1,\cdots,2q,\\ 
	  a_{mn}^{\alpha-2q}& \text{if } \alpha=2q+1,\cdots, 3q,\\ 
 	  b_{mn}^{\alpha-3q}& \text{if } \alpha=3q+1,\cdots, 4q.\\ 
    \end{array}
\right.
\end{eqnarray}
The partial sum (\ref{sum1}) can be performed by noticing that only products involving
an even number of $\sn{mn}$ give a non-zero contribution. We also define $\coefk{k}=\sum_{\sn{}=-S}^{S}\sn{}^{2k}\exp(\Delta\sn{}^2)$ and the ordered products
$q_{mn}^{(k)}=\sum_{\alpha_1<\alpha_2<\cdots<\alpha_k}c_{mn}^{\alpha_1}c_{mn}^{\alpha_2}
\cdots c_{mn}^{\alpha_k}$ with $q_{mn}^{(0)}\equiv 1$.
With these notations, it is easy to show that the partial sum (\ref{sum1}) gives the 
commuting objects
\bb \nn
W_{mn}=\sum_{k=0}^{2q}\coefk{k}q_{mn}^{(2k)},
\ee
with $q_{mn}^{(4q)}=c_{mn}^1\cdots c_{mn}^{4q}$ the term of highest degree in Grassmann
variables. Finally the fermionic representation of the partition function 
reads $\PF=\Cst^{2L^2}\TR_{\{a,\bar{a},b,\bar{b}\}}\prod_{mn}W_{mn}$.
This is the exact fermionic representation of the partition function for any given spin-$S$ model. \\
In some cases, the weights $W_{mn}$ can be easily exponentiated. For the Ising model, the argument of the exponential is purely quadratic in the $c_{mn}^{\alpha}$'s and therefore the partition function can be written as a determinant \cite{Plechko2002}. In the Blume-Capel model, the argument of the exponential is a polynomial of degree 8 in Grassmann variables since there are 8 independent variables $(4q=8)$ \cite{Clusel08}. In general we expect the argument to be at most a polynomial of degree $4q$ in these variables, which can sometimes be reduced by partial integrations. Except for the case $q=1$ the partition function is not solvable.\\
In the thermodynamical limit  $L\rightarrow \infty$, however, we expect to be able to identify from the effective theory a massive and pure kinetic contributions in the infrared region where the continuous momenta ${\bold k}$ are small. The condition of criticality is determined usually by the vanishing mass $m$ of the effective theory. For example, in the Dirac or Majorana representation of the Ising model, the free energy, which is the integral over the Brillouin zone of momentum-dependent quantities $\ln(m^2+{\bold k}^2)$, is singular at $m=0$. The action is determined by the exponentiation of the $W_{mn}$
quantities, depending on the $4q$ Grassmannian fields, and is made of a $local$ part, containing all the local interactions, including 2-fermion, 3-fermion etc.. interactions at a given site, and a $kinetic$ part, containing all the terms involving space derivatives of different orders. It is difficult to obtain the full fermionic action with all the $kinetic$ terms in the general case. In this paper, we will neglect the latter, assuming that their contribution by renormalization to the mass is negligible near the critical point. This is true for the Ising model where the $kinetic$ part is purely quadratic in Grassmann variables and does not renormalize the mass. Space symmetries of these derivative terms also could prevent any renormalization.
Then we will show that the contribution to the partition function coming from the $local$ part only can be computed exactly, which is not a quadratic action but a polynomial of degree $4q$, the number of variables involved, and this defines a bare mass. We need for this to define first the formal derivatives of Grassmann variables \cite{Plechko99}, for example: $\partial_x a_{mn}=a_{mn}-a_{m-1n}$ and $\partial_y a_{mn}=a_{mn}-a_{mn-1}$. 
Then the $c$'s coefficients can be expressed in term of these derivatives such as $c_{mn}^1= x_q(\bar{a}_{mn}^q-\partial_x\bar{a}_{mn}^q)$. In the limit of large
$L$ and in the Fourier space, the first order derivatives account in the action for a small contribution in momenta ${\bold k}=2\pi (m,n)/L$, with $m,n\ll L$ positive integers, when amplitudes $|{\bold k}|$ become small. In this infrared regime, we assume here that we can neglect the derivatives : $c_{mn}^1\simeq x_q\bar{a}_{mn}^q,\cdots,c_{mn}^{q}\simeq x_1\bar{a}_{mn}^{1}$ and
$c_{mn}^{q+1}\simeq x_q\bar{b}_{mn}^q,\cdots,c_{mn}^{2q}\simeq x_1\bar{b}_{mn}^{1}$ : the weights $W_{mn}$ are then all decoupled, and the following bare mass $m_S$ can be defined:
\bb\label{mass}
\frac{\mass_S}{\Cst^2}=\int\Big [
\prod_{\alpha=1}^q
d\bar{a}_{mn}^{\alpha}da_{mn}^{\alpha}
d\bar{b}_{mn}^{\alpha}db_{mn}^{\alpha}
e^{a_{mn}^{\alpha}\bar{a}_{mn}^{\alpha}+b_{mn}^{\alpha}\bar{b}_{mn}^{\alpha}}
\Big ] W_{mn}.
\ee
 The different integrals in Eq. (\ref{mass}) can be evaluated exactly by noticing for example that the arguments of the exponential 
$b_{mn}^{\alpha}\bar{b}_{mn}^{\alpha}$ can be combined with a $a_{mn}^{\alpha}(x_{\alpha}\bar{a}_{mn}^{\alpha})$ that appears
in some of the $q^{(2k)}$ products to give a contribution $x_{\alpha}$. Indeed using 
the Grassmann integration rules $\int da.a=1$ and $\int da.1=0$, we can write
\bb\nn
\int d\bar{a}_{mn}^{\alpha}da_{mn}^{\alpha}
d\bar{b}_{mn}^{\alpha}db_{mn}^{\alpha}
e^{a_{mn}^{\alpha}\bar{a}_{mn}^{\alpha}+b_{mn}^{\alpha}\bar{b}_{mn}^{\alpha}} a_{mn}^{\alpha}(x_{\alpha}\bar{a}_{mn}^{\alpha}) =
x_{\alpha}.
\ee
Since the $q^{(2k)}$ are ordered, there are also signs to take into
account and coming from moving the variables $c^{\alpha}_{mn}$ before integration. We obtain after some combinatorial algebra
\begin{eqnarray}
\label{resmass}
\mass_S=\sum_{k=0}^{2q}\coefk{k}R_{k},
\end{eqnarray}
where we have defined the following quantities $R_0=u_0^2$, 
\begin{eqnarray} 
\label{sumR}
R_k=\sum_{l=0}^{k}\pol_l\pol_{k-l}\sgn{l,k-l},
\end{eqnarray}
and $\sgn{k,l}=1$ if $k$ and $l$ are both even, and $\sgn{k,l}=-1$ otherwise.
We now apply this result to different cases. For the Ising model ($S=1/2$)
we obtain $\mass_{1/2}=2e^{\Delta/4}[1-\sh(K/2)],$ which vanishes at the normalized Ising critical temperature $t_c=T_c/S^2=2.269\,185$, independent of $\Delta_0$.  For the Blume-Capel model ($S=1$) we find $\mass_1=1+2e^{\Delta}[1-\sh(2K)]$ and for $S=3/2$:
\bb
\label{m32}
\mass_{3/2}=2e^{\Delta/4}[1-\sh(K/2)]+2e^{9\Delta/4}[1-\sh(9K/2)].
\ee.\\

\begin{table}
 \caption{\label{table1}Critical temperatures at $\Delta_0=0$}
 \begin{ruledtabular}
 \begin{tabular}{c|ccccccc|c}
 Spin S & $S=1/2$ & $S=1$ & $S=3/2$ & $S=2$\\
\hline
$q$   & 1 & 2 & 5 & 6
\\
$t_c$ & $2.269\,185$ & $1.673\,971$ & $1.456\,694$ & $1.337\,812$
\\
Refs. & $2.269$ \cite{Butera03} & $1.689$ \cite{Fox73}, $1.695$ \cite{Silva02} & $1.461$ \cite{xavier98,Butera03,grandi04} & $1.336$ \cite{Butera03} 
\\
 & & $1.694$ \cite{Butera03}, $1.681$ \cite{xavier98} & 
\\
\hline
 Spin S & $S=5/2$ &  $S=3$ & $S\rightarrow\infty$\\
\hline
$q$ &  11 & 12 & $\infty$
\\
$t_c$ & $1.262\,542$ & $1.210\,534$ & $0.925\,148$
\\
Refs. &  $1.257$ \cite{Butera03} & $1.203$ \cite{Butera03} & $0.915$ \cite{Butera03,Bial00}
 \end{tabular}
 \end{ruledtabular}
 \end{table}
For general spin $S$, we can show that
\bb
\mass_{S}=\sum_{\sigma=-S}^{S}e^{\Delta\sigma^2}[1-\sh(2\sigma^2 K)].
\label{genmass}
\ee
Equation (\ref{genmass}) gives the expression of the bare mass of a general spin-$S$ system, taking into account all possible local fermion-fermion interactions. To go further we propose to extract some physical information from the previous result in 
different cases. In particular, for the simplest non-integrable case (Blume-Capel model $S=1$), it is also possible to write explicitly the fermionic action and to check Eq. (\ref{genmass}) \cite{Clusel08}. Note however that even in this case a vanishing mass is a necessary but not sufficient condition to have a critical point : kinetic terms of higher order that appear in the effective fermionic action can change the nature of the singularity. For the Blume-Capel model for instance the critical line terminates at a tricritical point which can not be predicted by the mass alone \cite{Clusel08}. \\
Tabulated values of $t_c$ at $\Delta_0=0$ are given in Table \ref{table1} for several $S$, and compared with numerical results (Monte-Carlo simulations, high-temperature expansions) given in the literature. In general, the agreement is good.
\begin{figure}
\includegraphics[width=\columnwidth]{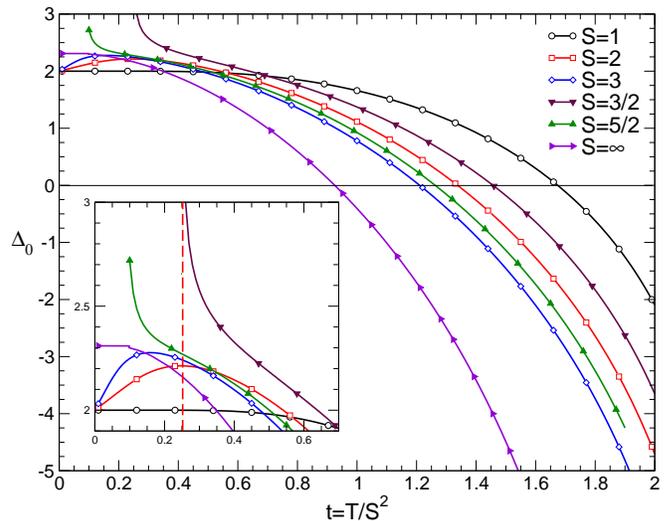}
\caption{\label{CriticalL} Critical line for different values of $S=1,2,3$, $S=3/2,5/2$
and $S=\infty$. In the latter case, the critical temperature at $\Delta_0=0$ is $t_c\simeq0.925\,148$, and at $t=0$ the curve reaches the solution $\Delta_0=4/\sqrt{3}$. The inset is a zoom on the region $\Delta_0>2$. The critical curve for $S=3/2$ has an asymptote (dotted vertical line) at $t=2/9\log(1+\sqrt{2})\simeq0.252\,132$.}
\end{figure}
For integer values of $S$ (Blume-Capel model), the critical line goes from the Ising critical value $t_c=2/\log(1+\sqrt{2})$ when $\Delta_0\rightarrow -\infty$ to the terminating point $(t_c=0,\Delta_0=2)$ continuously.  
For half-integer values of $S$, there exists in general an asymptote in the $(t=T/S^2,\Delta_0)$ plane. Indeed, for $S=3/2$, Eq. (\ref{m32}) predicts the solution
\bb
\Delta_0=-\frac{9t}{8}\log\Big [
-\frac{1-\sh(2/9t)}{1-\sh(2/t)}
\Big ],
\ee
which is bounded by $t_c=2/9\log(1+\sqrt{2})\simeq0.252\,131$ below which there is no
second-order critical line (see Fig.\ref{CriticalL}).
In the large integer $S$ limit, the model defined in Eq. (\ref{ham1}) is described by a continuous variable $-1<x_{mn}=\sn{mn}/S<1$ (continuous Ising model). We can obtain the limiting value of the mass, Eq. (\ref{genmass}) becoming:
\bb
\mass_{S\gg 1}\simeq S\sqrt{2t}\int_0^{\sqrt{2/t}}\dd x\; e^{-\Delta_0x^2/2}\Big [
1-\sh(x^2) \Big ].
\ee
We observe that the rescaled mass $\mass_S/S$ vanishes when $\Delta_0=0$ at $t_c\simeq 0.925\,148$, in fairly agreement with numerical works \cite{Butera03,Bial00}, and 
there is a non trivial solution at $t=0$ which is simply given by $\Delta_0=4/\sqrt{3} \simeq 2.309\,401$. Contrary to the finite $S$ models, the critical
field predicted here is not equal to $\Delta_0=2$, which gives the location of the 
first-order transition at zero temperature, but takes a slightly larger value. \\

By using Grassmann algebra to represent the partition function of spin-$S$ Ising Hamiltonians on 2D square lattices as a fermionic theory, we were able to obtain the exact expression for the bare mass of the action, including all the possible local fermionic interactions. This result gives at least precise though approximate location of second-order critical points in the $(T,\Delta_0)$ plane. This scheme and its main consequences, formulas (\ref{resmass}) and (\ref{sumR}), are general enough to be applied in a variety of classical Hamiltonians with next-nearest neighbor interactions and crystal field like potentials, and possibly Potts-like models as well, with a suitable choice of polynomial representation of the Boltzmann weights. 

\bibliography{ref}

\end{document}